
\magnification=\magstep1
\bigskip
\bigskip
\centerline{\bf PREFERRED OBSERVABLES, PREDICTABILITY, CLASSICALITY,}
\centerline{\bf AND THE ENVIRONMENT-INDUCED DECOHERENCE}
\medskip
\bigskip
\medskip
\centerline{Wojciech H. Zurek}
\bigskip
\bigskip
\centerline{Theoretical Astrophysics}
\centerline{T-6, MS B288}
\centerline{Los Alamos National Laboratory}
\centerline{Los Alamos, NM 87545}
\bigskip
\bigskip
\centerline{ABSTRACT}
{\narrower\smallskip\noindent Selection of the preferred classical set of
states in the process of
decoherence -- so important for cosmological considerations --
is discussed with an emphasis on the role of information loss and entropy.
{\it Persistence of correlations} between the observables of two systems (for
instance, a record and a state of a system evolved from the initial conditions
described by that record) in the presence of the environment is used
to define classical behavior. From the viewpoint of an observer (or any system
capable of maintaining records) {\it predictability} is a measure of such
persistence. {\it Predictability sieve} -- a procedure which employs both the
statistical and algorithmic entropies to systematically explore all of the
Hilbert space of an open system in order to eliminate the majority of the
unpredictable and non-classical states and to locate the islands of
predictability including the preferred {\it pointer basis} is proposed.
Predictably evolving states of decohering systems along with the time-ordered
sequences of records of their evolution define the effectively classical
branches of the universal wavefunction in the context of the ``Many  Worlds
Interpretation". The relation between the consistent histories approach and the
preferred basis is considered. It is demonstrated that histories of sequences
of
events corresponding to projections onto the states of the pointer basis are
consistent.
\smallskip}
\bigskip
\noindent{\bf 1. Introduction}
\medskip
The content of this paper is not a transcript of my presentation given in
course of the meeting: Proceedings contributions rarely are. However, in the
case of this paper the difference between them may be greater than usual.
The reason for the discrepancy is simple. Much of what I said is already
contained in papers (see, for example, my recent {\it Physics Today} article
(Zurek 1991) and references therein). While it would be reasonably easy to
produce a ``conference version" of this contribution, such an exercise would
be of little use in the present context. Nevertheless, I have decided to
include in the first part of this paper much of the same introductory material
already familiar to these interested in decoherence and in the problem of
transition from quantum to classical, but with emphasis on issues which are
often glossed over in the available literature on the subject, and which were
covered in the introductory part of my presentation in a manner more closely
resembling (Zurek 1991). The second part of the paper ventures into more
advanced issues which considered only briefly during the talk, but are the real
focus of attention here. In a sense, the present paper, while it is
self-contained, may be also regarded by reader as a ``commentary" on the
material discussed in the earlier publications. Its union with my 1991 paper
would be a more faithful record of my conference presentation, and a
familiarity
with that publication -- while not absolutely essential -- could prove helpful.
\medskip
{\it Decoherence} is a process which -- through the interaction of the system
with external degrees of freedom often referred to as the {\it environment} --
singles out a preferred set of states sometimes called {\it the pointer basis}.
This emergence of the preferred basis occurs through the ``negative selection"
process ({\it environment-induced superselection}) which, in effect, precludes
all but a small subset of the conceivable states in the Hilbert space of the
system from behaving in an effectively classical, predictable manner. Here I
intend to expand and enhance these past discussions in several directions:
\medskip
\itemitem (i) I shall emphasize the relation between the interpretational
problems of quantum theory and the existence, within the Universe, of the
distinct entities usually referred to as ``systems". Their presence, I will
argue, is indispensable for the formulation  of the demand of classicality.
This will justify a similar split of the Universe into the degrees of freedom
which are of direct interest to the observer -- a ``system of interest" --
and the remaining degrees of freedom known as the environment. It is therefore
not too surprising that the problem (in particular, the measurement problem,
but also, more generally, the problem of the emergence of classicality within
the quantum Universe) which cannot be posed without
recognizing systems cannot be solved without acknowledging that they exist.
\medskip
\itemitem (ii) I shall discuss the motivation for the concept of the preferred
set of ``classical" states in the Hilbert space of an open system
-- the pointer basis -- and show how it can be objectively implemented by
recognizing the role of the interaction between the system and the environment.
It will be noted that what matters is not a unique definition of the absolutely
preferred set of states, but, rather, a criterion for eliminating the vast
majority of the ``wrong", nonclassical states.
\medskip
\itemitem (iii) I shall consider the relation between the states which
habitually diagonalize (or nearly diagonalize) the density matrix of an open
system -- states which can be found on or near the diagonal after a decoherence
time has elapsed irregardless of what was the initial state -- and the
preferred
states. It will be pointed out that diagonality alone is only a symptom -- and
not a cause -- of the effective classicality of the preferred states. And, as
all symptoms, it has to be used with caution in diagnosing its origin: Causes
of
diagonality may, on occasion, differ from those relevant for the dynamics of
the
process of  decoherence (for example, ``accidental" diagonality in any basis
may be a result of complete ignorance of an observer) and the quantum to
classical transition that decoherence precipitates.
\medskip
\itemitem (iv) Therefore, I shall regard the ability to preserve correlations
between the records maintained by the observer and the evolved observables of
open systems as a defining feature of the preferred set of the to-be-classical
states, and generalize it to the situations where the observer is not present.
This definition, employed already in the early discussions of effective
classicality (Zurek, 1981), can be formulated in terms of information-theoretic
measures of the persistence of correlations, which leads one in turn to
consider
the role of {\it predictability} in the definition of the preferred basis.
\medskip
\itemitem(v) I shall analyse the relation of the decoherence approach
with the consistent histories interpretation. In particular, I shall show that
the histories expressed in terms of the events corresponding to the projections
onto the pointer basis states are consistent. Thus, a succesful decoherence
resulting in a stable pointer basis implies consistency. By contrast,
consistency alone does not constrain the histories enough to eliminate even
very non-classical branches of the state vector. Moreover, consistency
conditions single out not only the initial moment in which all the histories
begin, but also the final time instant on on a more or less equal footing.
Furthermore, in the absence of a fixed set of pointer projectors, consistent
histories cannot be expressed in terms of a ``global" set of alternative
events,
which could be then used for all histories at a certain instant of time, but,
rather, must be formulated in terms of ``local" events which are themselves
dependent on the past events in the history of the system, and which do not
apply to all of the  alternatives coexisting at a certain instant. Unless the
records are a part of the physical Universe, this is bound to introduce an
additional element of subjectivity into the consistent histories
interpretation.
\medskip
\itemitem(vi) When decoherence is succesful and does yield a preferred basis,
the branches of the universal wavefunction can thus be traced out by
investigating sequences of preferred, predictable sets of states singled out
by the environment - induced superselection. In this setting the issue of the
``collapse of the wavepacket" -- formulated as a question of accessibility,
through the existing records of the past events and the ability to use such
records to predict a future course of events -- is posed and discussed.
\medskip
I shall ``iterate" these themes throughout the course of the
paper, returning to most of them more than once. Consequently, the paper shall
cover the issues listed above only approximately in the order in which they
were itemized in this introduction. Sections 2 -- 5 cover the familiar ground
but do it with a new emphasis. Sections 6 -- 10 venture into a new territory
and should be of particular interest to these working on the subject.
\bigskip
\noindent {\bf 2. Motivation for Classicality, or ``No Systems - No Problem"}
\medskip
The measurement problem -- the most prominent example of the interpretational
difficulties one is faced with when dealing with the quantum -- disappears when
the apparatus - system combination is regarded as a single indivisible quantum
object. The same is true not just in course of measurements, but in general
-- the problems with the correspondence between quantum physics and the
familiar
everyday classical reality cannot be even posed when we refuse to acknowledge
the division of the Universe into separate entities.
\medskip
The reason for this crucial remark is simple. The Schr\"odinger equation;
$$i \hbar {d \over dt} \vert \Phi > = H \vert \Phi >  \ , \eqno (2.1)$$
is deterministic. Therefore, given the initial state, it predicts, with
certainty, the subsequent evolution of the state vector of any really isolated
entity, including the {\it joint} state of the apparatus interacting with the
measured system. Thus, in accord with Eq. (2.1), the initial state vector;
$$\vert \Phi_{\cal A \cal S}(0) > \ = \ | A_0 > | \phi_0 > \ = \ |A_0 >
\sum_i a_i | \sigma_i > \ , \eqno (2.2)$$
evolves into $\vert \Phi_{\cal A \cal S}(t) >$
in the combined Hilbert space $\cal H_{\cal A \cal S}$. There is no hint of the
interpretational problems in this statement until one recognizes that the
expected outcome of the measurement corresponds to a state which -- while
it contains the correlation between ${\cal A}$ and ${\cal S}$ -- also
acknowledges the ``right" of the apparatus to ``a state of its own".
By contrast, a typical form of $\vert \Phi_{\cal A \cal S} (t) >$ is;
$$\vert \Phi_{\cal A \cal S}(t) > = {\Sigma_i} a_i \vert A_i > \vert
\sigma_i > \ , \eqno (2.3)$$
where more than one $a_i(t)$ are non-zero. Above, $\{ \vert A_i >\}$ and $\{
\vert \sigma_i > \}$ belong to separate Hilbert spaces of the apparatus and of
the system, $\cal H_{\cal A}$ and $\cal H_{\cal S}$, such that
$$\cal H_{\cal A \cal S} = \cal H_{\cal A} \otimes \cal H_{\cal S}. \eqno
(2.4)$$

Let me emphasize at this point the distinction between quantum and classical
measurements by adopting a somewhat Dirac - like notation to discuss the
classical analog of a measurement - like evolution. The initial states of the
system ${\tt S}$ and of the apparatus ${\tt A}$ can be denoted by $|{\tt
s}_0\}$
and $| {\tt A}_0 \} $. When these states are ``pure" (completely known), and
the
evolution is classical, the final result of the measurement-like  co-evolution
is another ``pure" state: Reversible evolutions preserve purity in both quantum
and classical context, but with a final state which is simpler
classically than for a typical quantum case:
$$ | \Upsilon^{AS}_t \} = | {\tt A}_t \} | {\tt s}_t \}  \ . \eqno (2.5) $$
The key difference between Eq. (2.3) and Eq. (2.5) is the absence of the
superposition of the multiple alternatives corresponding to the possible
outcomes. And even when the classical initial states are known incompletely
(as is usually the case in the measurement situations)
the initial state of the system is
$$ \sum p_i | {\tt s}_i \} \ , \eqno (2.6) $$
where $p_i$ are the probabilities of the various states. The final state of the
interacting objects must be also described by a
classical probability distribution
$$\sum_i p_i |\Upsilon^{AS}_i \} = \sum_i p_i | {\tt A}_i \} | {\tt s}_i \}  \
{}.
\eqno(2.7)$$
Nevertheless, and in contrast with the quantum case, the range of the
final states -- pointer positions of the apparatus --  will continuously narrow
down as the range of the initial states of the system is being restricted. This
simultaneous constriction of the uncertainty about the initial state of the
system and about the outcome -- the final state of the apparatus -- shall be
referred to below as the {\it complete information limit}.
\medskip
Its consequence -- the expectation that a complete knowledge of the state
of a classical system implies an unlimited ability to predict an outcome of
every conceivable measurement -- is the cornerstone of our classical intuition.
This expectation is violated
by quantum theory: Complete information limit in the sense described above
does not exist in the quantum case. There, even in the case of the complete
information about each of the two objects (i.e., the system and the
apparatus) separately, one is still typically forced into a situation
represented by Eq. (2.3) in which -- following the measurement -- neither
the apparatus nor the system have ``a state of their own". Indeed, for every
pure state of the quantum system there exists a corresponding ``relative" state
of the apparatus (see Everett, 1957). In fact, a correlation between the states
of the apparatus and a measured system is, at this stage, analogous to
the nonseparable correlation between the two particles in the Einstein -
Podolsky - Rosen ``paradox" (Zurek, 1981).
\medskip
A closely related symptom of the quantum nature arises from the superposition
principle. That is, pure quantum states can be combined into another pure
state.
There is no equivalent way of combining pure classical states. The only
situation when their combinations are considered is inevitably tied to
probability distributions and a loss of information. Thus, formally, classical
states denoted above by $| \cdot \}$ correspond not to the vectors in the
Hilbert space of the system, but, rather, to the appropriate projection
operators, i. e.:
$$ | \cdot \} \Longleftrightarrow | \cdot >< \cdot | \eqno (2.8) $$
Such projection operators can be combined into probability distributions in
both
quantum and classical cases. A pure classical state -- a point in the phase
space -- is of course an extreme but convenient idealization. A more realistic
and useful concept corresponding to a probability distribution -- a patch in
the
phase space -- would be then described by the density matrix with the form
given
by Eq. (2.6).
\medskip
In a ``hybrid" notation the measurement carried out by a classical apparatus
on a quantum system would then involve a transition:
$$ |A_0 \} | \phi_0 > < \phi_0| \Longrightarrow \sum_i p_i | A_i \} |\sigma_i >
< \sigma_i | \eqno (2.9) $$
The difference between Eq. (2.3) and the right hand side of Eq. (2.9) is clear:
Classicality of the apparatus prevents the combined object from existing in
a superposition of states.
\medskip
The need for a random ``reduction of the state vector" arises as a result
of the demand -- based on the desire to reconcile consequences of Eq. (2.1)
(i.e., the linear superposition, Eq. (2.3)) with the expectation based on the
familiar experience (expressed above by Eqs. (2.5), (2.7) and (2.9)) that
only one of the outcomes should actually happen (or, at least, appear to
happen)
in the ``real world."
\medskip
There is, of course, a special set of circumstances in which the above
distinction between quantum and classical disappears, and no reduction of the
state vector following the interaction is required. This unusual situation
arises when -- already before the measurement -- the to-be-measured quantum
system was in an eigenstate (say, $| \sigma_k > $) or in a mixture of the
eigenstates of the measured observable. In the case of a single pure state
the resulting state of the apparatus - system pair following the interaction
between them is given by Eq. (2.3), but with only one non-zero coefficient,
$|a_i| ^2 =  \delta_{ik}$.
Hence, the distinction between Eq. (2.3) and Eq. (2.5) disappears. An obvious
generalization of this effectively classical case for a mixture results in an
appropriate probability distribution over the possible outcomes of the
measurement. Then the correlation between the system and the apparatus has a
purely classical character. Consequently, this situation is {\it operationally}
indistinguishable from the case when a classical system in
a state initially unknown to the observer is measured by a classical apparatus.
\medskip
These considerations motivate the decoherence approach -- an attempt to resolve
the apparent conflict between predictions of the Schr\"odinger equation and
perception of the classical reality. If the operationally significant
difference
between the quantum and classical characteristics of the observables can be
made
to continuously disappear, then maybe the macroscopic objects we encounter
are ``maintained" in the appropriate mixtures by the dynamical evolution
of the quantum Universe. Moreover, one can show that purely unitary evolution
of an isolated system can never accomplish this goal. Therefore, it is natural
to enquire whether one can accomplish it by ``opening" the system and allowing
it to interact with the environment. The study of this and related questions
defines the decoherence approach to the transition between quantum and
classical.
\bigskip
\noindent{\bf 3. Operational Goals of Decoherence}
\medskip
The criterion for success of the decoherence programme must be purely
operational. The Schr\"odinger equation and the superposition principle are at
its foundation. One must concede at the outset that they will not be violated
in
principle. Therefore, the appearance of the violation will have its origin in
the fact that the records made and accessed by observers are subject to
limitations, since they are governed by quantum laws with specific -- rather
than arbitrary -- interactions and since their records are not isolated from
the
environment. One important criterion -- to be discussed in more detail in the
next section -- refers to the fact that idealized classical measurements do
not change the state of the classical system. That is, for example, a complete
information limit -- the classically motivated idea that one can keep acquiring
information without changing the state of the measured object -- should be
an excellent approximation in the ``everyday" classical context. The
recognition
of the role of decoherence and the environment allows one to show how this
can be the case in a quantum Universe.
\medskip
An objection to the above programme is sometimes heard that -- in essence --
the Universe as a whole is still a single entity with no ``outside"
environment,
and, therefore, any resolution
involving its division into systems is unacceptable. While I am convinced that
much needs to be done in order to understand what constitutes a ``system", I
have also little doubt that the key aspects of the resolution proposed here are
largely independent of the details of the answer to this question.
As we have argued in the previous section, without
the assumption of a preexisting division of the Universe into individual
systems
the requirement that they have a right to their own states cannot be even
formulated: The state of a perfectly isolated fragment of the Universe -- or,
for that matter, of the quantum Universe as a whole -- would evolve forever
deterministically in accord with Eq. (2.1). The issue of the ``classicality"
of its individual components -- systems -- cannot be even posed. The
effectively
stochastic aspect of the quantum evolution (the ``reduction of the state
vector") is needed only to restore the right of a macroscopic (but ultimately
quantum) system to be in its own state.
\medskip
Quantum theory allows one to consider the wave function of the Universe.
However, entities existing within the quantum Universe -- in particular us, the
observers -- are obviously incapable of measuring that wave function
``from the outside". Thus, the only sensible subject of considerations aimed at
the interpretation of quantum theory -- that is, at establishing correspondence
between the quantum formalism and the events perceived by us -- is the relation
between the universal state vector and the states of memory (records) of
somewhat special systems -- such as observers -- which are, of necessity,
perceiving
that Universe from within. It is the inability to appreciate the consequences
of
this rather simple but fundamental observation that has led to such desperate
measures as the search for an alternative to quantum physics. One of the goals
of this paper is to convince the reader that such desperate steps are
unwarranted.
\medskip
One might be concerned that the appeal to systems and a frequent mention of
measurement in the above paragraphs heralds considerations with a character
which is subjective and may be even inexcusably ``anthropocentric", so that the
conclusions could not apply to places devoid of observers or to epochs in the
prehistory of the Universe. A few remarks on this subject are in order. The
rules we shall arrive at to determine the classical preferred basis will refer
only to the basic physical ingredients of the model (that is, systems, their
states, and the hamiltonians that generate their evolution) and we shall employ
nothing but unitary evolution. These rules will apply wherever these basic
ingredients are present, regardless of whether an observer is there to reap the
benefits of the emerging classicality. Thus, the role of ``an observer" will be
very limited: It will supply little more than a motivation for considering the
question of classicality.
\medskip
Having said all of the above, one must also admit that there may be
circumstances in which some of the ingredients appropriate for the Universe we
are familiar with may be difficult or even impossible to find. For example, in
the early Universe it might be difficult to separate out time (and, hence, to
talk about evolution). While such difficulties have to be acknowledged their
importance should not be exaggerated: The state vector of the Universe will
evolve in precisely the same way irregardless of whether we know what
observables could have been considered as ``effectively classical" in such
circumstances. Therefore, later on in course of its evolution, when one can
define time and discern the other ingredients relevant for the definition of
classicality, the question of whether they have been present early on or what
was or was not classical ``at the time when there was no time" will have
absolutely no observable consequences.
\medskip
This last remark deserves one additional caveat: It is conceivable (and has
been
even suggested, for example by Penrose (1989)) that the quantum theory which
must be applied to the Universe as a whole will differ from the quantum
mechanics we have become accustomed to in some fundamental manner (that is, for
instance, by violating the principle of superposition for phenomena involving
gravitation in an essential way). While, at present, there is no way to settle
this question, one can nevertheless convincingly argue that the coupling with
the gravitational degrees of freedom is simply too weak in many of the
situations involving ordinary, everyday quantum to classical transitions (for
example, the blackening of a grain of photographic emulsion caused by a single
photon) to play a role in the emergence of the classicality of everyday
experience.
\bigskip
\noindent {\bf 4. Insensitivity to Measurements and Classical Reality.}
\medskip
The most conspicuous feature of quantum systems is their sensitivity to
measurements. Measurement of a quantum system automatically results in a
{\it preparation}: It forces the system into one of the eigenstates of the
measured observable, which, from that instant on, acts as a new initial
condition. The idealized version of this quantum ``fact of life" is known as
the
projection postulate. This feature of quantum systems has probably
contributed more than anything else to the perception of a conflict between
the {\it objective reality}
of the states of classical objects and the elusive {\it subjective nature}
of quantum states which -- it would appear -- can be molded into a nearly
arbitrary form depending solely on the way in which they are being measured.
Moreover, the Schr\"odinger equation will typically evolve the system from the
prepared initial state (which is usually simple) into a complicated
superposition
of eigenstates of the initially measured observables. This is especially true
when an interaction between several systems is involved. Then the initial state
prepared by a typical measurement involves certain properties of each of the
separately measured systems, but quantum evolution results in
quantum correlations -- in states which are qualitatively similar to the one
given by Eq. (2.3). Thus, when the measurement of the same set of observables
is repeated, a new ``reduction of the wavepacket" and a consequent restarting
of the Schr\"odinger evolution with the new initial condition is necessary.
\medskip
By contrast, evolutions of classical systems are -- in our experience --
independent of the measurements we, the observers, carry out. A measurement
can, of course, increase our knowledge of the state of the system -- and,
thus, increase our ability to predict its future behavior -- but it should have
no effect on the state of the classical system {\it per se}. This expectation
is
based on a simple fact: Predictions made for a classical system on the basis of
the past measurements are not (or, at least, need  not be) invalidated by an
intermediate measurement. In classical physics, there is no need to
reinitialize
evolution, a measurement does not lead to the preparation, only to an update of
the records. This {\it insensitivity to measurements} appears to be a defining
feature of the classical domain.
\medskip
For a quantum system complete insensitivity to measurements will occur only
under special circumstances -- when the density matrix of the system commutes
with the measured observable already before the measurement was carried out.
In general, the state of the system will be influenced to some degree by the
measurements. To quantify the degree to which the state of the system is
altered by the measurement one can compare its density matrix before and after
the measurement (idealized here in the way introduced by von Neumann (1932)):
$$ \rho_{\rm after} \ = \ \sum_i P_i \rho_{\rm before} P_i \eqno(4.1) $$
The projection operators $P_i$ correspond to the various possible outcomes.
A generalization to the situation when a sequence of measurements is carried
out is straightforward:
$$ \rho_{\rm after} \ = \ \sum_{i,j, \dots , n} P_n \dots P_j P_i \rho_{\rm
before}
P_i P_j \dots P_n \eqno (4.2) $$
Above, pairs of projection operators will correspond to distinct observables
measured at consecutive instants of time. Evolution inbetween the measurements
could be also incorporated into this discussion.
\medskip
It is assumed above that even though the measurements have happened, their
results are not accessible. Therefore, the statistical entropy;
$$ h (\rho) \ = \ -Tr \rho \ln \rho \eqno(4.3)$$
can only increase;
$$ h (\rho_{\rm after} ) \geq h(\rho_{\rm before} ) \ . \eqno(4.4) $$
The size of this increase can be used as a measure of the sensitivity to a
measurement. Unfortunately, it is not an objective measure: It depends on how
much was known about the system initially. For example, if the initial state is
completely unknown so that $ \rho_{\rm before} \sim {\bf 1} $, there can be no
increase of entropy. It is therefore useful to constrain the problem somewhat
by assuming, for example, that the initial entropy $h (\rho_{\rm before} )$
is fixed and smaller than the maximal entropy. We shall pursue this strategy
below, in course of the discussion of the predictability in section 6.
\medskip
The role of decoherence is to force macroscopic systems into mixtures of states
-- approximate eigenstates of the same set of effectively classical
observables.
Density matrices of this form will be then insensitive to measurements of these
observables (that is, $\rho_{\rm after} \approx \rho_{\rm before}$). Such
measurements will be effectively classical in the sense described near the end
of Section 2, since the final, correlated state of the system and the apparatus
will be faithfully described by Eqs. (2.7) and (2.9). No additional reduction
of
the state vector will be needed, as the projection operators $P_i$ defining
the eigenspaces of the measured observable commute with the states which
invariably (that is, regardless of the initial form of the state of the system)
appear on the diagonal of the density matrix describing the system.
\medskip
This feat is accomplished  by the coupling to the environment -- which, as is
well known, results in noise, and, therefore, in a degraded predictability.
However, it also leads to a specification of what are the preferred observables
-- what set of states constitutes the preferred classical ``pointer basis"
which
can be measured relatively safely, with a minimal loss of previously acquired
information.  This is because the rate at which the  predictability is degraded
is crucially dependent on the state of the system. The initial state is in turn
determined by what was measured. We shall show that the differences in the
rates
at which predictability is lost are sufficiently dramatic that only for certain
selections of the sets of states -- and of the corresponding observables -- can
one expect to develop the idealized but very useful approximation of
classical reality.
\medskip
Another, complementary way of viewing the process of decoherence is to
recognize
that the environment acts, in effect, as an observer continuously monitoring
certain preferred observables which are selected mainly by the system -
environment interaction hamiltonians. Other physical systems which perform
observations (such as the ``real" observers) will gain little predictive power
from a measurement which prepares the system in a state which does not nearly
coincide with the states already monitored by the environment. Thus, the demand
for predictability forces observers to conform and measure what is already
being
monitored by the environment. Indeed, in nearly all familiar circumstances,
observers gain information by ``bleeding off" small amounts of the information
which has been already disseminated into the environment (for example by
intercepting a fraction of photons which have scattered off the object one is
looking at while the rest of the photons act effectively as an environment).
In such a case, the observer must be prepared to work with the observables
premeasured by the environment. Moreover, even if it was somehow possible to
make a measurement of some other very different set of observables with the
eigenstates which are given by exotic superpositions of the states monitored by
the environment, records of such observables would become useless on a
timescale
on which the environment is monitoring the preferred states. For, the
``measurements" carried out by the environment would invalidate such data on a
{\it decoherence timescale} -- that is, well before the predictions about the
future measurements made on its basis could be verified.
\medskip
In what follows we shall therefore search for the sets of observables as well
as the corresponding sets of states of open quantum systems
which offer optimal predictability of their own future values. This requirement
embodies a recognition of the fact that our measuring equipment (including our
senses) used in keeping track of the familiar reality is relatively
time-independent, and that the only measurements we are capable of refer to
small fragments of the Universe -- individual systems. We shall be considering
sequences of measurements carried out on ensembles of identical open quantum
systems. We shall return to the discussion of the
practical implementation of this programme in Sections 6 and 7, having
introduced -- in the next section -- a simple and tractable model in which
decoherence can be conveniently studied.
\bigskip
\noindent {\bf 5. Environment-Induced Decoherence in Quantum Brownian Motion.}
\medskip
The separation of the Universe into subsystems, indispensable in stating
the problem, is also instrumental in pointing out a way to its resolution.
Our experience proves that classical systems -- such as an apparatus --
have a right to individual states. Moreover, these states are not some
arbitrary states admissible in $\cal H_{\cal A}$ which can be selected on a
whim by an external observer (as would be the case for an ordinary quantum
system) but, rather, appear to be ``stable" in that the state in which $\cal
A$ is eventually found is always one of the ``eigenstates" of the same
``menu" of options -- i.e. the apparatus has a preferred ``pointer observable."
Thus, even though the correlated system - apparatus state vector $\vert
\Phi_{\cal A \cal S} (t) >$, Eq. (2.3), can be in principle rewritten using any
(complete) basis in $\cal H_{\cal A}$ (Zurek, 1981; 1991), this formal
application of the superposition principle and the resulting analogy with
nonseparable EPR correlations is not reflected in ``the familiar reality". The
familiar macroscopic systems tend to be well localized with respect to the
usual classical observables -- such as position, energy, etc.
\medskip
Here we shall see how this tendency towards localization, as well as the
choice of the preferred observables can be accounted for by the process of
decoherence: A classical system (such as an apparatus) will be continuously
interacting with its environment. Consequently, its density matrix will obey an
irreversible master equation, valid whenever the environment is sufficiently
large to have a Poincar\'e recurrence time much longer than any other relevant
timescale in the problem -- in particular, much longer than the timescale over
which one may attempt to predict its behavior or to verify its classicality.
\medskip
A useful example of such a master equation can be derived -- under certain
reasonable, although not quite realistic circumstances -- for a particle of
mass $m$ moving in a potential $V(x)$, and interacting with the external
scalar field (the environment) through the hamiltonian
$${\cal H}_{int} = \epsilon x {\dot \varphi} (0). \eqno (5.1)$$
In the high-temperature limit the reduced density matrix of the particle will
obey the equation (written below in the position representation):
$${d \over dt} \rho(x, x') = -{i \over \hbar} [ H_R, \rho] -
\gamma (x - x') ({\partial \over \partial x} - {\partial \over \partial x'})
\rho - {\eta k_B T \over \hbar^2} (x - x')^2 \rho. \eqno (5.2)$$
Above $H_R$ is the hamiltonian of the particle with $V_R(x)$ renormalized by
the
interaction with the field, $\eta$ is the viscosity coefficient $(\eta =
{\epsilon^2 \over 2}, \gamma = {\eta \over 2m}$), and $T$ is the temperature of
the field. Similar equations have been derived under a variety of assumptions
by, for example, Caldeira and Leggett (1983; 1985), Joos and Zeh (1985),
Unruh and Zurek (1989), and Hu, Paz and Zhang (1992).
\medskip
The preferred basis emerges through the process of ``negative selection": In
the
``classical" regime associated with $\hbar$ small relative to the effective
action the last term of Eq. (5.2) dominates and induces the following evolution
of the density matrix:
$$\rho(x, x'; t) = \rho(x, x';0) \exp(-{\eta k_B T \over \hbar^2} (x - x')^2
t).\eqno (5.3)$$
Thus, superpositions in position will disappear on the decoherence timescale
of order:
$$\tau_D = \gamma^{-1} (\lambda_T/\Delta x)^2 \eqno (5.4)$$
where $\gamma^{-1}$ is the relaxation rate
$$< \dot v > = - \gamma < v > \eqno (5.5)$$
This observation\footnote*{See Zurek (1984) for the first discussion of this
{\it decoherence timescale}; Zurek (1991) from the Wigner distribution function
perspective; Hartle (1991) for a nice perspective based on effective action;
Paz (1992) for a quick assessment of the role of the spectral density in the
environment on the effectiveness of decoherence and Paz, Habib and Zurek (1992)
for a more complete assessment of this important issue as well as for a
discussion of the accuracy of the high temperature approximation, Eq. (5.2).}
marked an  important shift from the idealized models of decoherence in the
context of quantum measurements to the study of more relistic models better
suited to the discussion of classicality in the ``everyday" context. The
conclusion is simple: Only relatively localized states will survive interaction
with the environment. Spread-out superpositions of localized wavepackets will
be
rapidly destroyed as they will quickly evolve into mixtures of localized
states.
\medskip
This special role of the position can be traced to the form of the
hamiltonian of interaction between the system and the environment, Eq. (5.1):
The dependence on $x$ implies a sensitivity to position, which is, in effect,
continuously monitored by the environment. In general, the observable
$\Lambda$ which will commute with $H_{int}$;
$$[H_{int}, \Lambda] = 0 \ , \eqno (5.6)$$
will play a key role in determining the preferred basis: Superpositions of
its eigenspaces will be unstable in the presence of the interaction with
the environment. Thus, the special role position plays in our description of
the
physical Universe can be traced (Zurek, 1986) to the form of the interaction
potentials between its subsystems (including the elementary particles and
atoms)
which is almost invariably position dependent (and, therefore,
commutes with the position observable).
\medskip
The result of such processes is the ``effective superselection rule," which
destroys superpositions of the eigenspaces of $\Lambda$, so that the density
matrix describing the open system will very quickly become approximately
co-diagonal with $\Lambda$. Thus, the classical system can be thought of --
on a timescale of several $\tau_D$ -- as a mixture (and not a superposition)
of the approximate eigenstates of position defined with the accuracy of the
corresponding $\Delta x$. This approximation becomes better when the self
hamiltonian of the system is less important when compared to $H_{int}$. Thus,
for example, overdamped systems have preferred states which are squeezed in
position, while weakly damped harmonic oscillator translates position into
momentum every quarter of its period and, consequently, favors still localized,
but much more symmetric coherent states. Moreover, for macroscopic masses and
separations the decoherence timescale is very small compared to the relaxation
time (Zurek, 1984 and 1991).
\medskip
The ability to describe the state of the system in terms of the probability
distribution of the same few variables is the essence of effective
classicality: It corresponds to the assertion that the system already has
its own state (one of the states of the preferred basis) which is not
necessarily known to the observer prior to the measurement, but is,
nevertheless, already quite definite. It is important to emphasize that while
this statement is, strictly speaking, incorrect (there is actually a very messy
entangled state vector including both the system and the environment) it
cannot be falsified by any feasible measurement. Thus, the quantum Universe
gives rise to an effectively classical domain defined through the operational
validity of the assertions concerning ``reality" of the states of the
observables it contains.
\bigskip
\noindent {\bf 6. Preferred States and the ``Predictability Sieve"}
\medskip
In the preceding section we have confirmed that -- at least for the model
presented above -- states localized in the familiar classical observable,
position, will be among these selected by the decoherence process and,
therefore, are effectively classical. Here I shall  reverse the procedure.
That is, I shall first formulate a test which will act, in effect, as a sieve,
a filter accepting certain states in the Hilbert space of the system and
rejecting others. The aim of this procedure will be to arrive at an algorithm
capable of singling out preferred sets of states {\it ab initio}, rather than
just confirming the suspicions about classicality of certain observables.
The guide in devising the appropriate procedure is a decade - old observation
(Zurek, 1981; 1982) that -- under some special assumptions -- the absolutely
preferred states remain completely unaffected by the environment. In the
absence
of these special conditions it is therefore natural to search for the states
which are least affected by the interaction with the external degrees of
freedom (Zurek, 1984; Unruh and Zurek, 1989). And a convenient measure
of the influence of the environment on the
state is -- as was already mentioned -- the ability to predict its future
evolution, to use the results of the past measurements as initial conditions.
\medskip
Let us consider an infinite ensemble of identical systems immersed in identical
environments. The aim of the test will be to find which initial states allow
for
optimal predictability. In order to settle this question, we shall imagine
testing {\it all} the states in the Hilbert space of the system ${\cal H_S}$.
To
this end, we shall prepare the system in every conceivable pure initial state,
let it evolve for a fixed amount of time, and then determine the resulting
final state -- which, because of the interaction with the environment, will be
nearly always given by a mixture.
\medskip
Entropy of the final density matrix $\rho_k$ which has evolved from the
initial state $|k>$;
$$ h_k \ = \ -Tr \rho_k \log \rho_k \eqno (6.1)$$
is a convenient measure of the loss of predictability. It will be different for
different initial states. One can now sort the list of all possible pure states
in order of the decreasing entropy $h_k$ of the final mixed state: At the head
will be the states for which the increase of entropy is the least,
followed by the states which are slightly less predictable, and so on, until
all of the pure states are assigned a place on the list.
\medskip
The best candidates for the classical ``preferred states" are, clearly, near
the
top of the list. One may, nevertheless, ask where should one put a cut
in the above list to define a border between the preferred classical set of
states and the non-classical remainder. The answer to this question is somewhat
arbitrary: Where exactly the quantum-classical border is erected is a
subjective
matter, to be decided by circumstances. What is, perhaps, more important is the
nature of the answer which emerges from this procedure aimed at selecting the
preferred set of states.
\medskip
A somewhat different in detail, but similar in the spirit procedure for sorting
the Hilbert space would start with the same initial list of the states and
evolve them until the same loss of predictability -- quantified by the entropy
of the final mixture -- has occurred. The list of states would be then sorted
according to the length of time involved in the predictability loss, with the
states which are predictable to the same degree for the longest time appearing
on the top of the list.
\medskip
It should be pointed out that the qualifying preferred states will generally
form (i) an overcomplete set, and (ii) they may be confined to a subspace of
${\cal H_S}$. These likely results of the application of the ``predictability
sieve" outlined above will have to be treated as a ``fact of life''.
In particular, overcompleteness seems to be a feature of many a ``nice" basis
--
for example, coherent states are overcomplete. Moreover, it is entirely
conceivable that classical states may not exist for a given system in all of
its
Hilbert space.
\medskip
It is reassuring to note that, in the situations in which an exact pointer
observable exists (that is, there is a nontrivial observable which commutes
with
the complete hamiltonian, $H + H_{int}$, of the system of interest; see Zurek,
1981-1983) its eigenstates can be immediately placed on top of
either of the predictability lists described above. The predictability sieve
selects them because they are completely predictable; they do not evolve at
all.
Hence, there is no corresponding increase of entropy.
\medskip
One key ingredient of the first of the predictability sieves outlined above
remains arbitrary: We have
not specified the interval of time over which the evolution is supposed to take
place. This aspect of the selection process will clearly influence the entropy
generated in the system. We shall insist that this time interval should remain
somewhat arbitrary. That is, the set of the selected preferred states for any
time $t$ much longer than the typical decoherence timescale $\tau_D$, but much
shorter than the timescale $\tau_{EQ}$ over which the system reaches
thermodynamic equilibrium with its environment should be substantially similar.
Similarly, for the second proposed sieve we have not specified the loss of
predictability (increase of entropy) which must be used to define the timescale
characterizing the location of the state on the list. Again, we expect that the
list should not be too sensitive to this parameter: Indeed, the two versions of
the sieve should yield similar sets of preferred states.
\medskip
The distinction between the preferred set of states and the rest of the Hilbert
space -- the contrast indicated by the predictability sieve outlined above --
should depend on the mass of the system or on the value of the Planck constant.
Thus, it should be possible to increase the contrast between the preferred set
of states and the other states which decohere much more rapidly by
investigating
the behavior of the entropy with varying $m$ and $\hbar$ in mathematical
models of these systems. It is easy to see how such a procedure will work in
the
case of quantum Brownian motion described in the preceding section. In
particular, reversible Newtonian dynamics can coexist with a very efficient
environment-induced superselection process in the appropriate limit
(Zurek, 1984; 1991): It is possible to increase mass,
decrease the Planck constant, and, at the same time, decrease the relaxation
rate to zero, so that the master equation Eq. (5.2) yields reversible Liouville
dynamics for the phase space probability distributions constructed from the
sets of localized states (such as the coherent states), but eliminates (by
degrading them into mixtures of the localized states) all other initial
conditions.
\medskip
A natural generalization of the predictability sieve can be applied to
mixtures.
Let us construct a list of all the mixtures which have the same initial entropy
$h_m(0) $ using either of the two prescriptions outlined above.
We can now consider their evolution and enquire about the entropy
of the mixed state after some time $t$. An ordered list of the initial mixtures
can be again constructed and the top of the list can be identified.
An interesting question can be now posed: Are the mixtures on top of the
initial mixture list diagonal in the pure states which are on top of the list
of preferred states? When the exact pointer pointer basis in the sense of the
early references on this subject (Zurek, 1981; 1982; 1983) exists, the
conjecture expressed above can be readily demonstrated: Mixtures of the pointer
basis eigenstates do not evolve at all. Therefore, their entropy remains the
same. Any other kind of mixture will become even more mixed because
superpositions of the pointer basis states decohere. Consequently, at least in
this case the most predictable mixtures are diagonal in the basis of the most
predictable states.
\medskip
In this preliminary discussion we have neglected  several issues which are
likely to play a role in the future developments of the predictability sieve
concept. In particular, we have ignored to enforce explicitly the requirement
(stated in Section 4) that the measurements which have prepared the set of the
candidate states should be suitable for the predictions of their own future
outcomes. (For the closed systems this criterion naturally picks out complete
sets of commuting observables corresponding to the conserved quantities.)
Moreover, we have used only the statistical entropy to assess  the cost of
the predictability loss.
\medskip
An addition of an algorithmic component might make for an even more interesting
criterion. That is, the place of the pure state $| \ell >$ on the list would be
based on the value of the {\it physical entropy} (Zurek, 1989), given by the
sum
of the physical entropy with the algorithmic information content $K(| \ell >
)$:
$$ s ( | \ell > ) \ = \ h_{\ell} + K( | \ell > ) \ . \eqno (6.2) $$
Algorithmic contribution would discourage overly complicated states.
\medskip
It is interesting to consider a possibility of using a similar predictability
criterion to justify the division of the Universe into separate systems.
The argument (which will be made more precise elsewhere) would start with
the list of all of the possible states in some composite Hilbert space,
perhaps even the Hilbert space associated with the Universe as a whole. The
procedure would then be largely similar to the simpler version of the sieve
outlined above, but with one important change: One will be allowed to vary
the definition of the systems -- that is, to test various projection operators,
such as the projections corresponding to averaging process -- to establish the
most predictable collective coordinates, and, simultaneously, to obtain the
preferred sets of states corresponding to such observables.
\bigskip
\noindent{\bf 7.  Predictability Sieve: An Example}
\medskip
A simple example of the dependence of the entropy production rate on the set of
preferred states is afforded by a model of a single harmonic oscillator
interacting with a heat bath we have analysed in Section 5. In this
section we shall carry out a restricted version of the more ambitious
and general ``predictability sieve" programme for this case. In its complete
version one would search for the set of states which minimize the entropy over
a finite time duration of the order of the dynamical timescale of the system.
Below we shall work with an instantaneous {\it rate} of change of {\it linear}
entropy. A more detailed analysis can be found elsewhere
(Zurek, Habib and Paz, 1992).
\medskip
Linear entropy
$$\varsigma ( \rho) \ = \ Tr (\rho  - \rho^2) \eqno(7.1) $$
is losely related to the ``real thing" through the expansion:
$$ h ( \rho ) \ = \ - Tr \rho \ln \rho \ \approx \ Tr \rho \{ ({\bf 1} - \rho)
+({\bf 1} - \rho )^2 + \cdots \ = \ \varsigma (\rho) + \cdots \eqno(7.2) $$
Moreover, it is in its own right a recognized measure of the purity of a state.
Consequently, in what follows I shall ``cut corners" and consider the quantity:
$$ \dot \varsigma \ = \ {d ~ Tr \rho^2 \over dt} \ = \ Tr ( \dot \rho \rho
+ \rho \dot \rho) \ \eqno(7.3) $$
for the initial pure states (that is, for all the density operators for which
$\rho^2 = \rho$). By contrast, the more ambitious programme would
deal with entropy increase over finite time of the order of the dynamical
timescale of the system.
\medskip
In the high temperature limit the equation which governs the evolution of the
reduced density operator can be written in the operator form as (Caldeira and
Leggett, 1983):
$$\dot \rho \ = \ {1 \over {i \hbar}} [ H_R, \rho] ~
+ ~ { \gamma \over {2 i \hbar}} [ \{ p, x \}, \rho ] ~
- ~ {{\eta k_B T} \over {\hbar^2}} [x, [x, \rho]] ~ - ~
{{i \gamma } \over \hbar} ( [x, \rho p] ~ - ~ [p, \rho x] ) \ . \eqno (7.4) $$
Above, square brackets and curly brackets indicate commutators and
anticommutators, respectively, while $H_R$ is the renormalized hamiltonian.
\medskip
Only the last two terms can be directly responsible for generation of linear
entropy; For, when the evolution of the density matrix is generated by
the term of the form  $\dot \rho \ = \ [ A, \rho], $ then the linear entropy
is unaffected.
The proof of the above assertion is simple: It suffices to substitute
in the expression for $\dot \varsigma$, Eq. (7.3), the above commutator form
for the  evolution generator of $\rho$. Then it is easy to compute:
$$ \dot \varsigma \ = \ Tr (A \rho^2 - \rho^2 A) $$
which is obviously zero by the cyclic property of trace.
\medskip
Consequently, after some algebra one obtains:
$$ \dot \varsigma \ = \ {4 \eta k_B T \over \hbar^2} Tr (\rho^2 x^2 - \rho x
\rho x)
\ - \ 2 \gamma Tr \rho^2 \ \eqno(7.5) $$
When the system is nearly reversible ($\gamma \approx 0$) the second term in
the
above equation is negligible. This is especially true in the high temperature
``macroscopic" limit (large mass, small Planck constant). It should be pointed
out that the initial ``decrease of entropy" predicted by this last term for
sufficiently localized wavepackets is unphysical and an artifact of the high
temperature approximation (Ambegaokar, private communication; Hu, Paz, and
Zhang, 1992). Nevertheless, at late times the existence of that term allows
for the eventual equilibrium with $\dot \varsigma = 0$.
\medskip
For the pure states the first term can be easily evaluated:
$$ \dot \varsigma \ = \
{{4 \eta k_B T } \over  \hbar^2 }
\cdot ( <x^2> \ - \ <x>^2) \ , \eqno (7.6)$$
where $<x>^2 = | < \varphi | x | \varphi> |^2 $ and $ <x^2> = < \varphi | x^2
| \varphi > $. This is an interesting result. We have just demonstrated that
{\it for pure states the rate of increase of linear entropy in quantum Brownian
motion is given by their dispersion in position, which is the preferred
observable singled out by the interaction hamiltonian}.
More generally, when the system is described by a density matrix so that
Eq. (7.5) must be used, it is easy to see that the density matrix which has
fewer superpositions of distinct values of $x$ will result in a smaller
entropy production. Thus, the special role of the observable responsible for
the
coupling with the environment is rather transparent in both of the above
equations.
\medskip
In view of the preceding considerations
on the role of predictability it is now easy to understand why localized states
are indeed favored and form the classical preferred basis. It should be
emphasized that this simple result cannot be used to infer that the preferred
states are simply eigestates of position. This is because the self-hamiltonian
-- while it does not contribute directly to the entropy production -- has an
important indirect effect. For a harmonic oscillator it rotates the state in
the phase space, so that the spread in momentum becomes ``translated" into the
spread in position and {\it vice versa}. As a result, linear entropy
produced over one oscillator period $\tau$ is given by:
$$ \varsigma_{\tau} \ = \   {{4 \eta k_B T } \over  \hbar^2 }
\int_0^{\tau} dt <\psi | ((x - <x>) \cos \omega t
+ (m \omega)^{-1} (p - <p>) \sin \omega t )^2 | \psi >$$
$$ = {{2 \eta k_B T } \over  \hbar^2 } (\Delta x_0^2 + {{\Delta p_0^2} \over
{m^2 \omega^2}} ) \ . \eqno(7.7) $$
Above, the dispersions are computed for the pure initial state $|\psi >$
and $\omega$ is the frequency of the oscillator.
\medskip
It is now not difficult to argue that the initial which minimizes linear
entropy
production for a harmonic oscillator must be (i) a minimum uncertainty
wavepacket, and (ii) its ``squeeze parameter" $s$ -- equal to the initial ratio
between its spread in position and the spread of the ground state -- must be
equal to unity. This can be confirmed by more detailed calculations, which
yield the value of $Tr \rho^2$ for the reduced density matrices which have
evolved from various initial states with different $s$ as a function of time.
Coherent states (minimum uncertainty wavepackets with $s  = 1$) are clearly
selected by the predictability sieve acting on a dynamical timescale of the
oscillator (see Fig. 1). This supplies a physical reason for the role they play
in harmonic motion, including its examples encountered in quantum optics.
\bigskip
\noindent{\bf 8. Insensitivity to Measurements and Consistent Histories}
\medskip
In 1984 Griffiths introduced a notion of history into the vocabulary used in
the discussion of quantum systems. A history is a time-ordered sequence of
properties represented by means of projection operators:
$$ \chi_{i,j,\dots,l} \ = \ P_i(t_i)E(t_i,t_j)P_j(t_j) \dots P_l(t_l)
E(t_l,t_0) \eqno (8.1) $$
where $t_i > t_j > \dots > t_l $, $P$'s stand for the projection operators
(assumed to be mutually orthogonal at each of the discrete instants at which
the corresponding properties are defined) and $E(t,t')$ evolve the system in
the time interval $(t,t')$. The set of histories is called {\it consistent}
when they do not interfere, that is, when the probabilities corresponding to
different histories can be added. Griffiths (1984) and Omn\`es (1990,1992)
(who emphasized the potential importance of the consistent histories approach
to the interpretation of quantum theory) have demonstrated
that a necessary and sufficient condition for the set of histories composed
of the initial condition and just two additional events to be consistent is
that
the projection operators defining the properties at the instants 0, 1, and 2,
when expressed in the Heisenberg picture, satisfy the algebraic relation;
$$ Re ~ Tr P_1 P_0 \bar P_1 P_2 \ = \ 0  \ , \eqno (8.2) $$
or, equivalently;
$$ Tr [P_1, [P_0, \bar P_1 ]] P_2 \ = \ 0 \ .  \eqno (8.3) $$
These equivalent consistency conditions must hold
for every of the projection operators in the sets defined at the initial,
intermediate, and final instants. Above, $\bar P_i = {\bf 1} - P_i$ projects on
the complement of the Hilbert space. Above, the explicit reference to the time
instants at which different projection operators act has been abbreviated to
a single index to emphasize that the time ordering is assumed to hold. There is
also a stronger consistency condition (Gell-Mann and Hartle, 1990)
with a physical motivation which draws on
the decoherence process we have discussed above: It is possible to show that
a sufficient condition for a family of histories to be consistent is the
vanishing of the off-diagonal elements of the {\it decoherence functional}:
$$ D(\chi, \chi') \ = \ Tr(P_2 P_1 P_0 P_1') \eqno (8.4)$$
Generalization to histories with more intermediate properties or with
the initial condition given by a density matrix is straightforward. One of the
appealing features of the Gell-Mann - Hartle approach is that it supplies
a convenient measure of the degree of inconsistency for the proposed sets of
histories.
\medskip
In all of the above consistency requirements above $P_0$ acts as an initial
condition and $P_2$ (or, more generally, the set of projection operators
corresponding to the ``last" instant) acts as a final condition. The two
sets of projection operators determine the initial and final sets of
alternatives for the system. (It is of course possible to start the system
in a mixture of different $P_0$'s.) This special role of the initial
and final sets of projection operators was emphasized
already by Griffiths (1984). It is particularly easy to appreciate in the
version of the consistent histories approach due to Gell-Mann and Hartle (1990;
1991; see also their contribution in this volume).
\medskip
Mathematically, the similarity between the role of the initial and final
projections in the formula for $D(\chi, \chi')$ follows from the cyclic
property of the trace. It is reflected in the special physical function of
these
two instants, which is usually defined by appealing to the physical content
of the situation to which the consistent  histories approach is applied.
For example, in the context originally considered by Griffiths the initial
condition and the set of final alternative properties were suggested by the
detectors contained inside the closed system under investigation (see
especially
section 3 of Griffiths, 1984). That is, the consistent histories approach is
capable of disposing of the outside intervention at the intermediate instants,
but not at the initial or final moments. Of course, in search for consistency
one can explore a variety of selections for the final set of projectors. Thus,
if there was just a single, unique set of consistent histories, one would be
still guaranteed to find it by employing the search procedure involving first
a selection of the preferred sets of initial and final projectors, and then
enquiring whether intermediate projectors consistent with these initial and
final conditions can be found. Unfortunately, consistent histories can be found
rather easily simply by evolving initial (or final) sets of projectors.
Moreover, these histories, while they are perfectly consistent, seem to have
little in common with the classical ``familiar reality". Indeed, the demand for
strict consistency will most likely have to be relaxed if the quantummechanical
histories are to be consistent with the reality we perceive. In this context,
where the demand for consistency is easily satisfied, and where the consistent
histories are crucially influenced by the initial and final selections, there
is
reason for concern with the ability of such ``outside" selections to unduly
influence the results.
\medskip
In the cosmological context the initial density matrix could be perhaps
supplied
by quantum cosmological considerations  but the final set of alternatives
is more difficult to justify. This distinction between the role of the
intermediate and final projections is worth emphasizing
especially because it is somewhat obscured by the notation which does not
distinguish between the special nature of the ``last" property and the
associated  set of the projection operators and the intermediate properties.
(The initial time instant is usually explicitly recognized as different by
notation, since density matrices are typically considered for the initial
condition in the majority of the papers on the subject (Omn\`es, 1992;
Gell-Mann
and Hartle, 1990).) Yet, the final set of options should not be -- especially
in the  cosmological context, where one is indeed dealing with the closed
system
-- determined ``from the outside". The question therefore arises: How can
one discuss consistency of histories without appealing to this final set
of ``special" events?
\medskip
Evolving density matrix can be, at any instant $t$, always written as:
$$ \rho (t) \ = \ \sum_{\cal K} \sum_{\cal K'} \chi({\cal K}) \rho(0)
\chi({\cal K'}) . \eqno (8.5) $$
Above $\chi ({\cal K}) $ is a collection of the projection operators with the
appropriate evolution operators sandwiched inbetween, and can be thought of
as a hypothetical history of the system. A composite index $ {\cal K}$
designates a specific time - ordered sequence. Projectors corresponding to
${\cal K}$ and ${\cal K'}$ differ, but the time instants or evolutions do not.
\medskip
Equation (8.5) is simply an identity, valid for arbitrary sets of orthocomplete
projection
operators defined at arbitrary intermediate instants $t_i,\ 0<t_i \leq t$.
There
is an obvious formal similarity between the expression for the decoherence
functional, Eq. (8.4), and the identity, Eq. (8.5). If one were to perform a
trace on the elements of $\rho(t)$ corresponding to pairs of different
histories
written in the above manner, one would immediately recover the expression
for the elements of the decoherence functional. The questions of consistency
of the sequences of events should be therefore at least partially
testable at the level of description afforded by the density matrix written in
this ``historical" notation. In particular, additivity of probabilities of the
distinct sequences of events corresponds simply to the {\it diagonality} of the
above expression for $\rho(t)$ in the set of considered histories. That is, the
set of histories $\{ \chi \}$ will be called {\it intrinsically consistent}
when
they satisfy the equality:
$$ \rho (t) \ = \ \bar \rho (t) \ , \eqno (8.6) $$
where:
$$\bar \rho(t) = \sum_{\cal K} \chi ( {\cal K}) \rho (0) \chi ({\cal K}) .
\eqno(8.7)$$
The physical sense of the equality between the double sum of Eq. (8.5) (which
is satisfied for every complete set of choices of the intermediate properties)
and the single sum immediately above (which can be satisfied for only very
special choices of $\chi$'s) is that the
interference terms between the properties defining the intrinsically consistent
sets of histories do not contribute to the final result;
$$ \sum_{{\cal K} \neq {\cal K}'} \chi ({\cal K}) \rho(0) \chi ({\cal K}') = 0
. \eqno (8.8) $$
An equivalent way of expressing the physical content of this equivalence
condition, Eqs. (8.6) or (8.8), is to use the terminology of Section 4 and note
that the system is insensitive to the measurement of the properties
defined by such intrinsically consistent histories.
An obvious way in which the above equation can be satisfied, and the only way
in which it can be satisfied exactly without demanding cancellations between
the
the off-diagonal terms, is to require that histories should be
expressed in terms of the properties which correspond to the operators
which commute with the density matrix. This simple remark (which, on
a formal level, implies the consistency condition, Eq. (8.4)) is, I believe,
essential in understanding the connection between the consistent histories
approach and the discussions based on the process of decoherence.
\medskip
An obvious exact solution suggested by the above considerations -- a set of
histories based on the evolved states which diagonalize the initial density
matrix -- does supply a formally acceptable set of consistent histories for
a closed system, but it is very unattractive in the cosmological, or
indeed, any practical context.
The resulting histories do not seem to have much in common with the familiar
classical evolutions we have grown accustomed to. It is clear that additional
constraints will need to be imposed, and the search for such constraints is
under way (Gell-Mann and Hartle, 1990; 1991; Omn\`es, 1992).
\medskip
In the decoherence approach presented in this paper the additional constraints
on the acceptable sets of consistent histories are supplied through insistence
on the observables which correspond to separate systems. This assumption
significantly constrains sets of the ``properties of interest" (see the
contribution of Albrecht to these proceedings as well as Albrecht (1992) for a
discussion of this point). Indeed, the restriction is so severe that one can
no longer hope for the exact probability sum rules to be satisfied at all
times.
Thus, instead of the ideal additivity of
the probabilities (or ideal diagonality of the density matrix in the set of
preferred pointer states) one must settle for approximations to the ideal which
violate it at a negligible level almost all the time, and which may violate
it at a significant level for a negligible fraction of time.
\medskip
Insensitivity of the system to measurements proposed in Section 4 as a
criterion
for classicality emerges as a useful concept unifying these two approaches --
the environment-induced decoherence and the consistent histories
interpretations. The condition of equality between the evolved density matrix,
with its double summation over the pairs of histories, and the restricted sum
over the diagonal elements only, Eq. (8.7), can be expressed
by noting that the measurements of the properties in terms of which the
consistent histories are written does not perturb the evolution of the system.
This suggests a physically appealing view of the consistent histories
approach in the context of the environment-induced decoherence: Measurements
of the preferred observable do not alter the evolution of the system which
is already monitored by the environment!
\bigskip
\noindent{\bf 9. Preferred Basis, Pointer Observables, and the Sum Rules}
\medskip
In the last section we have discussed a connection between the consistency of
histories and the notion of insensitivity of a quantum system to
measurements. Such insensitivity underlies much of our classical intuition.
It has served as an important conceptual input in the discussion of the
environment - induced superselection, decoherence, and the preferred pointer
basis. If the conceptual link between the foundations of the ``many histories
interpretation" and the preferred basis of an open system does indeed exist,
one should be able to exhibit a still more direct relationship between the
formalisms of these two approaches.
\medskip
Let us consider a system interacting with a large environment. In accord with
the preceding discussion, I shall focus on its reduced density matrix $\rho(t)
\equiv \rho_{\cal S}(t)$ which can be always computed by evolving the whole,
combined system  unitarily from the initial condition, and then tracing out the
environment. Throughout much of this section we shall also work under a more
restrictive assumption: We shall consider only situations in which the reduced
density  matrix of the system suffices to compute the decoherence functional.
It
should  be emphasized that this is not a trivial assumption, as the consistent
histories formalism was introduced for closed quantum systems and its extension
to the open quantum systems (required in essentially all practical
applications,
since macroscopic systems are nearly impossible to isolate from their
environments) leads to difficulties some of which have not yet been even
pointed out in the literature until now, let alone addressed.
\medskip
The nature of this demand is best appreciated by noting that (given a split
between the system and the environment and the usual tracing over the
environment degrees of freedom) we are demanding that the consistency of the
histories for the whole Universe with the events defined through
$P_k \otimes {\bf 1}_{\cal E}$ must be closely related to
the consistency of the histories with the events $P_k$ in the history of the
open system. Using the decoherence functional, the strongest form of such
equivalence could be expressed by the equality:
$$D_{\cal SE} \equiv Tr_{\cal SE} \{ {\bf 1}_{\cal E} \otimes P_l \ . \ . \ .
{\bf 1}_{\cal E}
\otimes P_i \ . \ . \ . \rho_{\cal SE} \ . \ . \ . P_i' \otimes {\bf 1}_{\cal
E} ... \}
\ = \  Tr_{\cal S} \{ P_l \ . \ . \ . P_i \ . \ . \ . \rho \ . \ . \ . P_i'  \
. \ . \ .\} \equiv D_{\cal S} \eqno (9.1)$$
Above, ``. . ." stand for the appropriate evolutions inbetween events (which is
unitary for ${\cal SE}$ considered jointly and non-unitary for ${\cal S}$
alone). While the above equation has the advantage of stating a requirement
with
brevity, one should keep in mind not only that the diagonality of the
decoherence functional is a more restrictive condition than the one necessary
to satisfy the sum rules, but also that exact diagonality is not to be expected
in the physically interesting situations. Moreover, when the approximate
consistency is enough (as would have to be almost always the case) the equality
above could be replaced by an approximate equality, or could hold with a very
good accuracy almost always, but be violated for a fraction of time which is
sufficiently small not to be worrisome. Finally, it is conceivable that the
approximate equality could hold only for a certain sets of projection
operators,
and not for others.  We shall not attempt to analyse these possibilities here.
For the time being, we only note that whenever the density matrix of the open
system obeys a master equation which is local in time (such as Eq. (5.2))
the reduced density matrix will suffice to express the decoherence functional.
These issues will be discussed in a separate paper (Paz and Zurek, 1992).
\medskip
To begin the discussion of the connection between the pointer basis and the
consistent histories approach I shall assume -- guided by the experience with
the environment - induced superselection -- that after a decoherence time the
density matrix of the system settles into a state which, irregardless of the
initial form of the state of the system ends up being nearly diagonal in the
preferred basis $P_i$.  I shall, for
the moment, demand exact diagonality and show that it implies exact
consistency.
\medskip
Let us first consider an initial state which is a projection operator $Q_o$,
and a historical property which corresponds to a pointer observable:
$$ \Lambda^i \  = \  \sum_i \lambda_i P_i \ , \eqno(9.2)$$
which commutes with the reduced density matrix:
$$ \rho_{Q_o}(t^i) \ = \ Tr_{\cal E} \rho_{\cal SE} (t^i) \ , \eqno(9.3)$$
at the first historical instant;
$$ [ \Lambda^i, \rho_{Q_o}(t_i) ] \ = \ 0 \ . \eqno(9.4) $$
Above, the lower case ``$i$" plays a double role: When in subscript, it is
a running index which can be summed over (as in Eq. (9.2)). When in the
superscript it is used to label the instant of time (i. e. $t^i, \ t^j \ , $
etc.) or a pointer observable ($\Lambda^i $) at that instant.
\medskip
Given Eq. (9.2), the density matrix can be written as:
$$ \rho_{Q_o} (t^i) \ = \ \sum_i p(i|o) \rho_{ii|o} \ = \ \sum_i P_i \rho_{Q_o}
(t^i) P_i \ . \eqno(9.5)$$
In other words, the system is insensitive to the measurement of the pointer
observable at that instant. Above, $p(i|o)=Tr_{\cal S} P_i \rho_{Q_o}(t^i) P_i$
is the conditional probability that a pointer projector $P_i$ would have been
measured at a time $t^i$ given that the system started its evolution in the
state $Q_o$ at the initial instant. The conditional density matrix;
$$\rho_{ii|o} = P_i \rho_{Q_o}(t^i) P_i /p(i|o)  \eqno(9.6)$$
which appears in Eq. (9.5) will, in general, contain additional information
about the
measurements of higher resolution which could have been carried out. It would
reduce to unity when $Tr P_i = 1$.
\medskip
Given this form of the reduced density matrix at $t^i$ it is not difficult to
evaluate the contribution of that instant to the decoherence functional for
histories including other properties $ \{P_k\} $ at the corresponding
instants $t^k$:
$$D_{Q_o}(\chi_{i,j,...l}, \chi_{i',j',...l}) = Tr \{ ... (P_i \rho_{Q_o}(t_i)
P_i') ...\}=p(i|o) D_{\rho_{ii|o}}(\chi_{j,...l}, \chi_{j',...l}) \delta_{ii'}
\eqno(9.7) $$
Above, the projection is followed by the (nonunitary) evolution which continues
until the next projection, and so on, until the final instant $t_l$. The
off-diagonal elements of the decoherence functional are always zero because of
our simplifying assumption of existence of a perfect pointer basis at each
``historical" instant.
\medskip
Such computation can be carried out step by step until finally one can write;
$$ D_{Q_o}(\chi, \chi) = p(l|k...j|i|o) ... p(j|i|o) p(i|o) \ , \eqno (9.8)$$
and;
$$ D_{Q_o}(\chi \neq \chi') = 0 \   \eqno (9.9) $$
for the case when pointer observables existed and were used to define
properties
at each historical time instant.
\medskip
When the initial state was mixed and of the form;
$$ \rho(t^o) = \sum_o p(Q_o) Q_o \ , \eqno(9.10) $$
the elements of the above formula for the decoherence functional would have
to be weighted by the probabilities of the initial states, so that;
$$ D_{\rho} = \sum_o p(Q_o) D_{Q_o} \ . \eqno(9.11) $$
Conditional probabilities in Eq. (9.8) can be simplified when the ``historical
process" is Markovian, so that the probability of the ``next" event depends
only on the preceding historical event. Delineating the exact set of conditions
which must be satisfied for this Markovian property to hold is of interest,
but,
again, beyond the scope of this paper: Markovian property of the underlying
dissipative dynamics is not enough. One can, however, see that the process
will be Markovian when the projections are onto rays in the Hilbert space.
By contrast, conditional probabilities will be determined to a large extent
by the properties in the more distant past when such events have had a better
resolution -- that is, if they projected onto smaller subspaces of the Hilbert
space of the system -- than the more recent ones.
\medskip
We have thus seen than when (i) the density matrix of the system alone is
enough to study consistency of histories and (ii) when the system evolves into
density matrix which is exactly diagonal in the same set of pointer states
each time historical properties are established, then the histories expressed
in
terms of the pointer projectors are exactly consistent. In the more realistic
cases the reduced density matrix will be only approximately diagonal in the
set of preferred states. It is therefore of obvious interest to enquire how
the deviations from diagonality translate into a partial loss of consistency.
\medskip
To study this case we suppose now that the reduced density matrix of the system
at a time $t^i$ does not satisfy Eq. (9.4), but, rather, is given by;
$$ \rho_{Q_o}(t^i) = \sum_i p(i|o) \rho_{ii|o} \ + \
\sum_{i \neq i'} P_i \rho_{Q_o} (t^i) P_i' \ . \eqno(9.12) $$
Below, we shall set $\varrho_{ii'|o} = P_i \rho_{Q_o} (t^i) P_i'.  $
Following the similar procedure as before one obtains (for the diagonal
contribution at the instant $t^i$) the same formula as before. By contrast, an
off-diagonal term also contributes;
$$ D_{Q_o} ( \chi_{i,j,...l}, \chi_{i',j',...l}) = D_{\varrho_{ii'|o}}
(\chi_{j,...l}, \chi_{j',...l}) \ . \eqno(9.13)$$
The new ``initial density matrices" for such off-diagonal terms are
non-hermitean. Nevertheless, they can be evolved
using -- for example -- the appropriate master equation and will contribute to
the off-diagonal terms of the decoherence functional. Moreover, the size of
their contribution to the decoherence functional can be computed by writing
them as a product of a hermitean density matrix and an annihilation operator;
$$ \varrho_{ii'} = q_{ii'|o} ~ \rho_{ii'|o} ~ \Pi_{ii'} \ , \eqno(9.14)$$
where;
$$ P_i = \Pi_{ii'} ~ P_i' ~ \Pi_{ii'}^{\dagger} \ , \eqno(9.15)$$
and $q(ii'|o)$ is the trace of this component of $\varrho_{ii'}$ which
remains after separating out $\Pi_{ii'}$.
\medskip
The simplest example of this procedure obtains in the case when the projectors
are one-dimensional. Then Eq. (9.14) simplifies and the off-diagonal element
of the decoherence functional can be evaluated by a formula similar to the
one for diagonal elements:
$$ D_{Q_o} ( \chi_{i,j,...l}, \chi_{i',j',...l}) = q(ii'|o) D_{\Pi_{ii'}}
(\chi_{j,...l}, \chi_{j',...l}) \ . \eqno(9.16) $$
Expressing the decoherence functional in this form allows one to estimate the
decrease of the potential for interference -- for the violations of the sum
rules -- from the relative sizes of the numerical coefficients such as $p(i|o)$
and $q(ii'|o)$, products of which establish the ``relative weight" of the
various on-diagonal and off-diagonal terms. Similar procedure applies also more
generally and leads to a formula:
$$ D_{Q_o} ( \chi_{i,j,...l}, \chi_{i',j',...l}) = q(ii'|o)
D_{\rho_{ii'|o}\Pi_{ii'}}  (\chi_{j,...l}, \chi_{j',...l}) \ . \eqno(9.17) $$
Equations (9.7), (9.8) and (9.17) -- representing individual steps in
evaluation of the
decoherence functional -- can be repeated to obtain ``chain formulas" of the
kind exhibited before in this section. We shall not go into such details which
are relatively straightforward conceptually but rather cumbersome notationally.
\medskip
These considerations establish a direct link between the process of decoherence
with its preferred observables, corresponding sets of states, etc.,
and the requirement of consistency as it is implemented by the decoherence
functional. It is now clear that -- given the assumptions stated early on in
this section -- the environment induced superselection implies consistency of
the histories expressed in terms of the pointer basis states. It should be
nevertheless emphasized that the issue of the applicability of the sum
rules and the consistency criteria address -- in spite of that link --
questions which differ, in their physical content, from these which are usually
posed and considered as a motivation for the environment - induced decoherence
and the resulting effective superselection.
\medskip
The key distinction arises simply from the rather limited goal of the
consistent histories programme -- the desire to ascribe probabilities
to quantum histories -- which can be very simply satisfied by letting
the histories follow the unitary evolution of the states which were on the
diagonal of the density matrix of the Universe at the initial instant.
This obvious answer has to be dismissed as irrelevant (Gell-Mann and Hartle,
1990) since it has nothing to do with the ``familiar reality". The obvious
question to ask is then whether the approximate compliance with the sum rules
(which have to be relaxed anyway to relate the histories to the ``familiar
reality") would not arise as a byproduct of a different set of more restrictive
requirements, such as these considered in the discussions of
environment-induced
decoherence. In the opinion of this author, the division of the Universe
into parts -- subsystems -- some of which are ``of interest" while other
act as ``the environment" is essential. It is required in stating the problem
of
interpretation of quantum theory. Moreover, it automatically rules out trivial
solutions of the kind discussed above, while, at the same time, allowing for
the consistency of histories expressed in terms of the preferred basis. In this
context the approximate validity of the sum rules which is of course an
important prerequisite for classicality, arises naturally as a consequence of
the opennes of the systems through the dynamics of the process of decoherence.
\bigskip
\noindent {\bf 10. Preferred Sets of States and Diagonality}
\medskip
Diagonality of the density matrix in some basis alone is not
enough to satisfy the physical criteria of classicality.
Pointer states should not be defined as the states which just happen to
diagonalize the reduced density matrix at a certain instant. What is crucial
(and sometimes forgotten in the discussions of these issues) is the requirement
of stability of the members of the preferred basis (expressed earlier in this
paper in terms of predictability). This was the reason why the preferred
observable was defined in the early papers on this subject through its
ability to retain correlations with other systems in spite of the contact
with the environment (Zurek, 1981). Hence, the early emphasis on the relation
of the pointer observable to the interaction hamiltonian and the analogy with
the monitoring of the ``nondemolition observables" (Caves et al., 1980).
\medskip
An additional somewhat more technical comment -- especially relevant in the
context of the many-histories approach -- might be in order. One might imagine
that the exact diagonality of the density matrix in some basis (rather than
existence of the stable pointer basis) might be employed to guarantee
consistency. After all, the property of the density matrix we have used earlier
in this section to establish consistency of the histories expressed in terms
of the pointer basis at a single instant, Eqs. (9.7) - (9.9) -- was based on
diagonality alone. This might be possible if one were dealing with a single
instant of time, but would come with a heavy price when the evolution of the
system is considered: The projection operators which diagonalize the reduced
density matrices which have evolved from certain ``events" in the history of
the system will typically depend on these past events! That is, the set of
projectors on the diagonal of $\rho_{Q_o}(t_i)$ will be different from the set
of projectors on the diagonal of $\rho_{Q_o'}(t_i)$ even when the initial
conditions -- say, $Q_o$ and $Q_o'$ -- initially commute, or even when they are
orthogonal. This is because the evolution of the open system does not preserve
the value of the scalar products. Thus, for instance, the implication;
$$ [ \rho(0), \rho'(0)] = 0 \Leftrightarrow [ \rho(t), \rho'(t) ] =0 \ ,
\eqno(10.1)$$
which is valid for unitary evolutions, does not hold for the evolution of open
systems. This can be established, for example, for the dissipative term
of the master equation (5.2).
\medskip
The implications of this phenomenon for both of the approaches considered here
remains to be explored. It seems, however, that consequences of such
environment-induced noncommutativity for the consistent histories approach
could be quite important: The standard formalism underlying
the many - histories interpretation of quantum mechanics requires an exhaustive
set of exclusive properties which can be defined for the system under study as
a whole (see papers by Griffiths, Omn\`es, Gell-Mann, and Hartle cited
earlier).
This requirement is perfectly reasonable when the system in question is
isolated
and does not include observers who decide what questions are being
asked -- what sets of projections operators will be selected -- as is
surely the case for this Universe (see Wheelers ``game of twenty questions"
in his 1979 paper).  It could be perhaps satisfied for macroscopic
quantum systems of the sort considered in the context of quantum measurements
for finite stretches of time. It is, however, difficult to imagine how it
could be satisfied for all the branches of the Universe and ``for eternity",
especially since the effective hamiltonians which decide what are the sensible
macroscopic observables are more often than not themselves determined by the
past events (including the dramatic symmetry breaking transitions which are
thought to have happened early on in the history of the Universe, as well as
much less dramatic, but also far less hypothetical transitions in macroscopic
systems). A similar comment can be made about the selection of the same
``historical" instants of time in a multifarious Universe composed of branches
which specifically differ in the instant when a certain event has occurred.
\medskip
The above remark leads one to suspect that such global (or, rather,
universal) orthogonal sets of events at reasonably simultaneous instants
of time cannot and satisfy the demands of consistency ``globally". Only
``local"
properties defined in a way which explicitly recognizes the events which have
occurred in the past history of the branch have a chance of fulfilling this
requirement. This suggests the interesting possibility that the consistent
history approach can be carried out with the same set of the event-defining
properties and for the same sequence of times only for a relatively local
neighborhood of some branch defined by the common set of essential ingredients
(which can be in turn traced to the events which contributed to their
presence).
Such branches would then also share a common set of records -- imprints of the
past events reflected in the ``present" states of physical systems.
It should be noted that while the standard formalism does not call for the
branch dependence, consistent histories framework can be generalized to
acommodate it. Indeed, both Omn\`es (1990) as well as Gell-Mann and Hartle
(1990) have mentioned such branch-dependent histories in their papers.
Environment-induced noncommutativity may force one to consider this possibility
even more seriously.
\bigskip
\noindent{\bf 11. Records, Correlations and Interpretations}
\medskip
The discussion near the end of the last section forces us to shift the focus
from the histories defined in terms of ``global" projection operators, abstract
``events", and equally abstract ``records" related to them and existing
``outside" of the Universal state vector to far more tangible ``local" events
which have shaped a specific evolution of ``our" branch, and which are
imprinted as the real, accessible records in the states of physical systems.
It also emphasizes that even an exceedingly ``nonperturbative" strategy of
trying to identify consistent causally connected sequences of events appears to
have inevitable and identifiable physical consequences.
\medskip
In a sense, we have come back full circle to the essence of the problem of
measurement understood here in a very non-anthropocentric manner, as a
correlation between the states of two systems, as an appearance of records
within -- rather than outside -- of the state vector of the Everett - like
Universe. Moreover, if our conclusions about the need to restrict the analysis
of consistent histories to local bundles of branches derived from the records
which are in principle accessible from within these branches are correct,
and the emphasis on the coarse - grainings which acknowledge a division of the
Universe into systems (existence of which will undoubtedly be branch -
dependent) is recognized, the difference between the two ways of investigating
classicality considered above begins to fade away.
\medskip
The central ingredient of the analysis is not just a state of the system, but,
rather, a correlation between states of two systems -- one of which can be
regarded as a record of the other. The states of the systems we perceive
as an essential part of classical reality are nothing but a ``shorthand" for
the existence of a reliable correlation between these states and the
corresponding records -- states of our memory (possibly extended by such
external devices as notebooks, RAM's, or -- on rare occasions -- papers in
conference proceedings). This ``shorthand" is sufficient only because the
states of physical system used as memories are quite stable, so that their
reliability (and effective classicality) can be taken for granted at least
in this context. Nevertheless, the only evidence for classicality comes not so
much from dealing with the bundles of histories which may or may not be
consistent, but, rather, from comparisons of these records, and from the
remarkable ability to use the old records as initial conditions which --
together with the regularities known as ``the laws of physics" -- allow one to
predict the content of the records which will be made in the future.
\medskip
If our discussion was really about the stability of correlations between the
records and the states (rather than just states) it seems appropriate to close
this paper with a brief restatement of the interpretation problem from such
``correlation" point of view. The entity we are concerned with is not just a
to - be - classical system ${\cal S}$ interacting with the environment, but,
rather, a combination of ${\cal S}$ with a memory ${\cal M}$, each immersed in
the environments ${\cal E}_{\cal S}$ and ${\cal E}_{\cal M}$.
The density matrix describing this combination must have -- after the
environment has intervened -- a form:
$$ \rho_{\cal MS} = Tr_{{\cal E}_{\cal S}} Tr_{{\cal E}_{\cal M}} \rho_{{\cal
MS}
{\cal E}_{\cal M} {\cal E}_{\cal S}}
= \sum p(i,j) P_i^{\cal M} P_j^{\cal S} \ , \eqno(11.1)$$
where $p(i,j)$ must be nearly a Kronecker delta in its indices;
$$ p(i,j) \approx \delta_{ij} \ , \eqno(11.2)$$
so that the states of memory $\{ P_i^{\cal M} \}$ can constitute a faithful
record of the states of the system.
\medskip
One might feel that these statements are rather transparent and do not
have to be illustrated by equations. Nevertheless, the two equations above
clarify some of the aspects of the environment - induced decoherence and
the emergence of the preferred sets of states which were emphasized from the
very beginning (Zurek, 1981; 1982) but seem to have been missed  by some of the
readers (and perhaps even an occasional writer) of the papers on the subject of
the transition from quantum to classical. The most obvious of these is the need
for stability of the correlations between the preferred sets of
states selected by the processes of decoherence occurring simultaneously in the
memory and in the system, and the fact that memory is a physical system
subjected to decoherence inasmuch as any other macroscopic and, therefore,
coupled to the environment degrees of freedom, open system.
\medskip
The second point is somewhat more subtle (and, consequently, it has proved to
be
more confusing): The form of the classical correlation expressed above by the
two equations precludes any possibility of defining preferred states in terms
of the density matrix of the system alone. The key question in discussing the
emergence of classical observables is not the set of states which are on the
diagonal after everything else (including the memory ${\cal M}$) is traced out,
but, rather, the set of states which can be faithfully recorded by the memory
-- so that the measurement repeated on a timescale long compared to the
decoherence time will yield a predictable result. Hence the focus on
predictability, on the form of the interaction hamiltonians, and on the minimum
entropy production, and the relative lack of concern with the states which
remain on the diagonal ``after everything but the system is traced out".
The diagonality of the density matrix -- emphasized so much in many of the
papers -- is a useful symptom of the existence of the preferred basis when it
occurs (to a good approximation) in the same basis. For, only existence of such
stable preferred set of states guarantees the ability of the to be classical
systems to preserve the classical amount of correlations. I shall not belabor
the discussion of this point any further: The reader may want to go back to the
discussion of the role of the pointer basis in preserving some of the
correlations in measurements (Zurek, 1981; 1982; 1983), as well as to the
earlier sections where some of the aspects of this issue were considered.
\medskip
Finally, the most subtle point clarified -- but only to some extent -- by the
above two equations is the issue of the perception of a unique reality. This
issue of the ``collapse of the wave packet" cannot be really avoided: After
all,
the role of an interpretation is to establish a correspondence between the
formalism of a theory describing a system and the results of the experiments
-- or, rather, the records of these results -- accessible to the observer.
And we perceive outcomes of measurements and other events originating at the
quantum level alternative by the alternative, rather than all of the
alternatives at once.
\medskip
An exhaustive answer to this question would undoubtedly have to involve a
model of ``consciousness", since what we are really asking concerns our
(observers) impression that ``we are conscious" of just one of the
alternatives. Such model of consciousness is presently not available.
However, we can investigate what sort of consequences about the collapse could
be arrived at from a rather more modest and plausible assumption that such
higher mental processes are based on information processing involving data
present in the accessible memory. ``To be conscious" must than mean that the
information processing unit has accessed the available data, that it has
performed logical operations including copying of certain selected records.
Such copying results in redundant records: The same information is now
contained in several memory cells. Therefore, questions concerning the
outcome of a particular observation can be confirmed (by comparing different
records) or verified (by comparing predictions based on the past records
with the outcomes of new observations) only within the branches.
\medskip
In a sense, the physical state of the information processing unit will be
significantly altered by the information-processing activity and its physical
consequences -- its very state is to some degree also a record. Information
processing which leads to ``becoming conscious" of different alternatives
will result in computers and other potentially conscious systems which differ
in their physical configuration, and which, because their state is a presently
accessible partial record of their past history, exist in different Universes,
or rather, in different branches of the same universal state vector. These
differences will be especially obvious when the machinery controlled by the
computer reacts to the different outcomes in different ways. A particularly
dramatic example of different fates of the machinery is afforded by the example
of Schr\"odingers Cat: The cat in the branches of the Universe in which it is
dead cannot be aware of any of the two alternatives because its
``machinery" was significantly modified by the outcome of the quantum event --
it became modified to the extent which makes it impossible for the cat to
process information, and, hence, to be aware of anything at all.
\medskip
I believe that similar, but (fortunately) much more subtle modification of both
the records and the identity -- the physical state -- of the recording and
information processing ``observer" (animated or otherwise) are the ultimate
cause of the ``collapse". This mechanism for the collapse is closely coupled
to the phenomenon of decoherence, which makes the changes of the machinery
irreversible. The selection of the alternatives arises because of the split
into the machinery (including the memory and the information processing unit)
and ``the rest of the Universe", through the environment induced
superselection.
The possible set of alternatives of both ``what I know" and ``what I am" is
then fixed by the same process and at the same time. An observer carries in
its state either an implicit or an explicit record of its branch. It should be
emphasized that these last few paragraphs are considerably more speculative
than
the rest of the paper, and should be treated as such by the reader.
\medskip
The interpretation that emerges from these considerations is obviously
consistent with the Everett's ``Many Worlds" point of view. It is supplemented
by a process -- decoherence -- which arises when the division of the Universe
into separate entities is recognized. Its key consequence -- emergence
of the preferred set of states which can exist for time long compared to the
decoherence timescale for a state randomly selected from the Hilbert space of
the system of interest -- is responsible for the selection of the individual
branches. As reported by an observer, whose memory and state becomes modified
each time a ``splitting" involving him as the system takes place, the
apparent collapses of the state vector occur in complete accord with the Bohr's
``Copenhagen Interpretation". The role of the decoherence is to establish
a boundary between quantum and classical. This boundary is in principle
movable,
but in practice largely immobilized by the irreversibility of the process of
decoherence (see Zeh, 1991) which is in turn closely tied to the number of the
degrees of freedom coupled to a macroscopic body. The equivalence between
``macroscopic" and ``classical" is then validated by the decoherence
considerations, but only as a consequence of the practical impossibility of
keeping objects which are macroscopic perfectly isolated.
\medskip
I would like to thank Robert Griffiths, Salman Habib, Jim Hartle and Juan
Pablo Paz for discussions and comments on the manuscript.
\bigskip
\noindent {\bf References}
\bigskip
\itemitem A. Albrecht (1992); ``Investigating Decoherence in a Simple System",
{\it Phys. Rev.}, to appear.
\itemitem A. O. Caldeira and A. J. Leggett (1983); ``Path Integral Approach to
Quantum Brownian Motion",
{\it Physica} {\bf 121 A}, 587.
\itemitem A. O. Caldeira and A. J. Leggett (1985); ``Influence of Damping on
Quantum Interference: An Exactly Soluble Model",
{\it Phys. Rev.} {\bf A 31}, 1057.
\itemitem C. M. Caves, K. S. Thorne, R. W. P. Drewer, V. D. Sandberg, M.
Zimmermann (1980); ``On a Measurement of a Weak Force Coupled to a Quantum
Harmonic Oscillator", {\it Rev. Mod. Phys.} {\bf 52}, 341.
\itemitem M. Gell-Mann and J. B. Hartle (1990); ``Quantum Mechanics in the
Light of Quantum Cosmology", p. 425 in {\it Complexity, Entropy, and the
Physics of Information}, W. H. Zurek, editor (Addison-Wesley).
\itemitem M. Gell-Mann and J. B. Hartle (1991); ``Alternative Decohering
Histories in Quantum Mechanics",  Caltech preprint CALT-68-1694.
\itemitem R. J. Griffiths (1984); ``Consistent Histories and the Interpretation
of Quantum Mechanics" {\it J. Stat. Phys.} {\bf 36}, 219.
\itemitem J. B. Hartle (1991); ``The Quantum Mechanics of Cosmology", in
{\it Quantum Cosmology and Baby Universes} S. Coleman, J. B. Hartle, T. Piran,
and S. Weinberg, eds. (World Scientific, Singapore).
\itemitem B. L. Hu, J. P. Paz, and Y. Zhang (1992); ``Quantum Brownian Motion
in a General Environment", {\it Phys. Rev.} {\bf D45}, 2843.
\itemitem E. Joos and H. D. Zeh (1985); ``The Emergence of Classical Properties
Through Interaction with the Environment",
{\it Z. Phys.} {\bf B 59}, 223.
\itemitem R. Omn\`es (1991) ``From Hilbert Space to Common Sense: A Synthesis
of Recent Progress in the Interpretation of Quantum Mechanics"
{\it Ann. Phys. (N. Y.)} {\bf 201}, 354.
\itemitem R. Omn\`es (1992) ``Consistent Interpretations of Quantum Mechanics";
{\it Rev. Mod. Phys.} {\bf 64}, 339.
\itemitem J. P. Paz, S. Habib, and W. H. Zurek (1992); ``Decoherence and the
Reduction of the Wavepacket", {\it Phys. Rev.} {\bf D}, submitted.
\itemitem J. P. Paz and W. H. Zurek (1992); ``Consistent Histories Approach to
Decoherence", Los Alamos preprint, in preparation.
\itemitem R. Penrose (1989); {\it The Emperors New Mind} (Oxford University
Press, Oxford).
\itemitem W. G. Unruh and W. H. Zurek (1989); ``Reduction of the Wavepacket in
Quantum Brownian Motion", {\it Phys. Rev.} {\bf D 40}, 1071.
\itemitem J. von Neumann (1932); {\it Mathematical Foundations of Quantum
Mechanics}, English translation by Robert T. Beyer (1955; Princeton University
Press, Princeton).
\itemitem J. A. Wheeler (1979); ``Frontiers of Time", in {\it Problems in the
Foundations of Physics, Proceedings of the International School of Physics
``Enrico Fermi"} (Course 72), N. Toraldo di Francia, ed. (North Holland,
Amsterdam).
\itemitem H. D. Zeh (1971); ``On the Irreversibility of Time and Observation
in Quantum Theory", in {\it Enrico Fermi School of Physics} {\bf IL},
B. d'Espagnat, ed. (Academic Press, New York).
\itemitem H. D. Zeh (1989) {\it The Physical Basis for the Direction of Time}
(Springer, Heidelberg).
\itemitem W. H. Zurek (1981); ``Pointer Basis of Quantum Apparatus: Into
What Mixture Does the Wave Packet Collapse?", {\it Phys. Rev.} {\bf D 24},
1516.
\itemitem W. H. Zurek (1982); ``Environment - Induced Superselection Rules",
{\it Phys. Rev.} {\bf D 26}, 1862.
\itemitem W. H. Zurek (1983); ``Information Transfer in Quantum Measurements:
Irreversibility and Amplification", p. 87 in {\it Quantum Optics, Experimental
Gravitation, and the Measurement Theory}, P. Meystre and M. O. Scully, eds.
(Plenum, New York).
\itemitem W. H. Zurek (1984); ``Reduction of the Wavepacket: How Long Does it
Take?", lecture at the NATO ASI {\it Frontiers of the Nonequilibrium
Statistical Mechanics} held in June 3 - 16 in Santa Fe, New Mexico, published
in
the proceedings, G. T. Moore and M. O. Scully, eds. (Plenum, New York, 1986).
\itemitem W. H. Zurek (1986); ``Reduction of the Wave Packet and Environment
- Induced Superselection", p. 89 in {\it New Techniques and Ideas in Quantum
Measurement Theory}, D. M. Greenberger, ed., (New York Academy of Sciences,
New York).
\itemitem W. H. Zurek (1989); ``Algorithmic Randomness and Physical Entropy",
{\it Phys. Rev.} {\bf A 40}, 4731.
\itemitem W. H. Zurek (1991); ``Decoherence and the Transition From Quantum to
Classical", {\it Physics Today}, {\bf 44} (10), 36.
\itemitem W. H. Zurek, S. Habib, and J. P. Paz (1992); ``Coherent States Via
Decoherence", Los Alamos preprint.
\bigskip
\vfill
\eject
\noindent{\bf Figure Caption}
\medskip
\noindent{Fig. 1. Predictability sieve for an underdamped harmonic oscillator.
The plot shows $Tr \rho^2 \sim 1/{\rm area} $ (which serves as a measure of
purity of the reduced density matrix $\rho$) for mixtures which have evolved
from the initial minimum uncertainty wavepackets with different squeeze
parameters $s$ in a damped harmonic oscillator with $\gamma / \omega =
10^{-4}$.
Coherent states ($s = 1$) which have the same spread in position and momentum,
when measured in the units natural for the harmonic oscillator, are clearly
favored as a preferred set of states by the predictability sieve. Thus, they
are expected to play the crucial role of the pointer basis in the transition
from quantum to classical.}
\bigskip
\vfill
\eject
\noindent{\bf Discussion}

\bigskip
\noindent{\bf Halliwell}: You have the nice result that the {\it off -
diagonal} terms
of the density matrix (i. e. parts representing interferences) become
exponentially suppressed in comparison with the on - diagonal terms. But why
does this mean that you have decoherence? Why is it the peaks of the density
matrix that matter? In short, what exactly is your definition of decoherence?
In the approach of Gell-Mann and Hartle, decoherence is defined rather
precisely by insisting that the probability sum rules are satisfied by the
probabilities of histories. These rules are trivialy satisfied, always, for
histories consisting of one moment of time. You seem to have only one moment
of time. How is your notion of decoherence reconciled with theirs?
\medskip
\noindent{\bf WHZ}: Very simply. I prefer to use the word ``consistency"
employed in the earlier work to describe
histories which satisfy probability sum rules (i. e. Griffiths, 1984)
and adopted by the others (Omn\`es, 1990; 1992, and references therein).

I feel that the term {\it decoherence} is best reserved for the {\it process}
which destroys quantum coherence when a system becomes correlated with the
``other" degrees of freedom, its environment (which can include also internal
degrees of freedom as would certainly be the case for the environment of a
pointer in a quantum apparatus). In spite of this ``conservative" attitude
to the nomenclature, I am partial to the ``decoherence functional" Gell-Mann
and
Hartle have employed to formulate their version of consistency conditions.
This is because much of the motivation for their work goes beyond simple
consistency, and in the direction which, when pursued for open systems, will
result in a picture quite similar to the one emerging from the recognition of
the role of the process of decoherence (see Sections 8 and 9 above).

Now as for your other queries; (i) Decoherence is a process, hence it happens
in
time. In recognition of this, the discussions of the effects of decoherence
have always focused on correlations and on the interaction hamiltonians used,
for example, to define preferred basis, rather than on the instantaneous state
of the density matrix, which can be always diagonalized.
(ii) Consistency rules are ``trivialy satisfied" at a single moment of time
because consistency conditions are always formulated in a manner which
effectively results in a ``collapse of the state vector" onto the set of
alternatives represented by projectors at both the beginning and the end of
each history. And applying a formalism which was developed to deal with
histories to an object defined at an instant is likely to be a bad idea --
and yield results which are either wrong or trivial.
(iii) Nevertheless, the effect of decoherence process can be studied at an
instant through the {\it insensitivity} of the system to measurements of the
preferred observables. This insensitivity can be related to the existence of
classical correlations and seems to be a natural criterion for the
{\it classicality} of the system. It sets in after a decoherence time
irregardless of the initial state of the system.

\bigskip

\noindent{\bf Griffiths}: The results you have discussed are very interesting,
and add to our understanding of quantum mechanics. On the other hand, I think
that they could be stated in a much more clearer way if you would use the
``grammar" of consistent histories. For example, in the latter interpretation
the choice of basis is not left to the ``environment"; it is a choice made by
the theoretician. In fact, you choose the basis that interests you and, from
the
consistent histories point of view, what you then did was to show that in this
choice of basis, and after a suitably short time, etc., the questions you want
to ask form a consistent history. But in order to think clearly about a
problem,
it is useful to use a clear language, rather than talking about ``fairies",
which is what one tends to do if one carries over the traditional language of
quantum measurement.
\medskip
\noindent{\bf WHZ}: I tend to believe that even though theoreticians have
more or less unlimited powers, they cannot settle questions such as the one
posed by Einstein in a letter to Born.\footnote*{The quote from a 1954 letter
from Albert Einstein to Max Born ``Let
$\Psi_1$ and $\Psi_2$  be solutions of the same Schr\"odinger equation. ...
When the system is a macrosystem and when $\Psi_1$ and $\Psi_2$ are `narrow'
with respect to the macrocoordinates, then in by far the greater number of
cases this is no longer true for $\Psi = \Psi_1 + \Psi_2$. Narrowness with
respect to macrocoordinates is not only {\it independent} of the principles
of quantum mechanics, but, moreover, incompatible with them..."  was on one
of the transparencies shown during the talk, quoted after the translation of E.
Joos from
{\it New Techniques and Ideas in Quantum Measurement Theory}, D. M.
Greenberger,
ed., (New York Acad. Sci., 1986)} Moreover, I find it difficult to dismiss
questions such as this as ``fairies". Furthermore, I feel rather strongly that
it is the openness of the macroscopic quantum systems which -- along with the
form of the interactions -- allows one to settle such issues without appealing
to anything outside of physics.

I agree with you on the anticipated relationship between the
consistent histories and decoherence, and on the need to establish
correspondence between these two formalisms. I believe that they are compatible
in more or less the manner you indicate, but I do not believe that they are
equivalent. In particular, consistency conditions alone are not as restrictive
as the process of decoherence tends to be. They can be, for example,
satisfied by violently non-classical histories which obtain from evolving
unitarily projection operators which initially diagonalize density matrix.
The relation between decoherence and consistency clearly requires further
study.

\medskip
\noindent{\bf Leibowitz}: I agree with much of Griffiths' comment although I
disagree with him about ``fairies". I believe that all questions are legitimate
and, as pointed out by Bell and fully analyzed in a recent paper by Durr,
Goldstein,
and Zangi ({\it J. Stat. Phys.,}, to appear) the Bohm theory which assigns
complete reality to particle positions gives a completely clear explanation
of the {\it non - relativistic} quantum mechanics which is free from both
problems of measurement and from ``having to talk to ones lawyer" before
answering some of the questions. Whether Bohm's theory is right I certainly do
not know, but it does have some advantages of clarity, which should be
considered when discussing ``difficulties" of quantum mechanics.

\noindent{\bf WHZ}: I am always concerned with whether the relativistic
version of the Bohm - de Broglie theory can even exist. (How could photons
follow anything but straight lines? Yet, in order to explain double - slit
experiments {a l\'a} Bohm - de Broglie one tends to require rather complicated
trajectories!) Moreover, having grown  accustomed to the probabilities
in quantum mechanics, I find the exact causality of particle trajectories
unappealing.

\bigskip
\noindent{\bf Omn\`es}: Concerning the basis in which the reduced density
operator becomes diagonal: For a macroscopic system this question is
linked with another one, which is to define the ``macrocoordinates" from first
principles. My own guess (relying on the work of the mathematician Charles
Fefferman) is that one will find diagonalization in terms of both position and
momentum, as expressed by quasiprojectors representing classical properties.
\medskip
\noindent{\bf WHZ}: I agree with your guess and I can base it on the special
role played by coherent states in a quantum harmonic oscillator (as it is
shown in Section 7 of the paper in more detail), but I believe that decoherence
is sufficient to understand similar spreads in position and momentum. The
symmetry between $x$ and $p$ is broken by the interaction hamiltonian (which
depends on position) but partially restored by the rotation of the
Wigner function corresponding to the state in the phase space by the
self-Hamiltonian. I am, however, not familiar with the work of Fefferman.
Therefore, I cannot comment on your motivation for your guess.

\bigskip

\noindent{\bf Bennett}: If the interaction is not exactly diagonal in position,
do the superpositions of position still decohere?
\medskip
\noindent{\bf WHZ}: This will depend on details of the interaction,
self-Hamiltonian, etc., as discussed in the paper and references. It is,
however, interesting to note that when the interaction hamiltonian is periodic
in position (and, therefore, diagonal in momentum) and the particles are
otherwise free (so that the self-Hamiltonian is also diagonal in momentum)
diagonalization in momentum (and delocalization in position) is expected for
the preferred states. Such situation occurs when electrons are traveling
through a regular lattice of a crystal, and the expectation about their
preferred states seems to be borne out by their behavior.

\bigskip

\noindent{\bf Hartle}: I would like to comment on the connection between the
notion of ``decoherence" as used by Wojciech in his talk and the concept of
``consistent histories" discussed by Bob Griffiths and Roland Omn\`es that was
called ``decoherent histories" in the work by Murray Gell-Mann and myself.
As Griffiths and Omn\`es discussed yesterday, probabilities can be assigned
to sets of alternative coarse - grained histories if, and only if, there is
nearly vanishing interference between the individual members of the set, that
is, if they ``decohere" or are ``consistent".

For those very special types of coarse grainings in which the fundamental
variables are divided into a set that is followed and a set that is ignored
it is possible to construct a reduced density matrix by tracing the full
density matrix over the ignored variables. Then one can show that typical
mechanisms that effect the decoherence of histories (as in the Feynman -
Vernon,
Caldiera - Leggett type of models) also cause the off - diagonal elements of
the reduced density matrix to evolve to small values (see, e. g., my lectures,
Hartle 1991). However, the approach of the reduced density matrix to
diagonality
should not be taken as a fundamental definition of decoherence. There are at
least two reasons: First, realistic coarse - grainings describing the
quasiclassical domain such as ranges of averaged densities of energy, momentum,
etc. do not correspond to a division of the fundamental variables into ones
that are distinguished by the coarse - graining and the others that are
ignored.
Thus, in general and realistic cases there is no precise notion of environment,
no basis associated with coarse - graining, and no reduced density matrix.

The second reason is that decoherence for histories consisting of alternatives
at a single moment of time is automatic. Decoherence is non-trivial only for
histories that involve several moments of time. Decoherence of histories,
therefore, cannot be defined at only one time like a reduced density matrix.
It is for these reasons that the general discussion of decoherence
(consistency)
is important.
\medskip
\noindent {\bf WHZ}: I have a strong feeling that most of the disagreement
between us that you outline in your comment is based on a difference in our
vocabularies. You seem to insist on using words ``decoherence" and
``consistency" interchangeably, as if they were synonymous. I believe that
such redundancy is wasteful, and that it is much more profitable to set up
a one - to - one correspondences between the words and concepts. I am happy to
follow the example of Griffiths and Omn\`es and use the term ``consistency" to
refer to the set of histories which satisfy the probability sum rules. I would
like, on the other hand, to reserve the term ``decoherence" to describe the
process which results in the loss of quantum coherence whenever (for example)
two systems such as a ``system of interest" and an ``environment" are becoming
correlated as a result of a dynamical interaction. This distinction seems to
be well established in the literature (see, in addition to Omn\`es, also
Albrecht, Conradi, DeWitt (in these proceedings) Griffiths (including a comment
after DeWitt's talk), Fukuyama, Habib, Halliwell, Hu, Kiefer, Laflamme,
Morikawa,
Padmanabhan, Paz, Unruh, and Zeh, and probably quite a few others). I see
little
gain and a tremendous potential for confusion in trying to change this existing
usage, especially since the term ``decoherence" is well - suited to replace
phrases such as ``loss of quantum coherence" or ``dephasing" which were used
to describe similar phenomena (although in more ``down to earth" contexts)
for a very long time. Thus, I am in almost complete
agreement with much of your comment providing that we agree to follow Griffiths
and Omn\`es and continue to use the word ``consistent" when referring to
histories which satisfy the sum rules. And since you indicate that the two
words can be used interchangeably, I assume that you will not object to this
proposal. Having done so (and after re-stating your comment with the
appropriate
substitutions) I find only a rather minor items which require further
clarification.

The procedure required to distinguish the system from the environment is one of
them. You seem to insist on such distinction appearing in the {\it fundamental}
variables (which, I assume, would probably take us well beyond electrons and
protons, to some version of the string theory). I do not believe that such
insistence
on fundamental variables is necessary or practical. Indeed, the questions
addressed in the context of transition from quantum to classical usually
concern variables such as the position and momentum of the center of mass,
which are not fundamental, but are nevertheless coupled to the rest of the
Hilbert space, with the environment consisting in part of the ``internal
variables".
For the hydrodynamic variables such split is accomplished by defining
appropriate projection operators which correspond to the averages (see, e. g.,
``Equilibrium and Nonequilibrium Statistical Mechanics" by Radu Balescu, as
well as papers by Zwanzig, Prigogine and R\'esibois, and others). Similarly,
in Josephson junctions (which motivated the work of Caldeira and Leggett on
the influence functional) the environment is ``internal" and at least one of
the macroscopic quantum observables (the current) is hydrodynamic in nature.

Finally, as to the question whether (1) decoherence and (2) consistency can be
studied at a single moment of time, I would like to note that: (i) Decoherence
is a process, which occurrs in time. Its effects can be seen by comparing the
initial density matrix with the final one. The corresponding instantaneous
rate of the entropy increase, as well as the discrepancy between the
instantaneous density matrix and the preferred set of states can be also
assessed at an instant, as discussed in Section 7 above. (ii) I agree with you
that it is pointless
to study consistency of histories which last an instant, since the first and
the last set of the
projection operators are ``special" and, in effect, enforce a ``collapse
of the wave packet" (see Sections 8 and 9 for details), and instantaneous
histories are rather degenerate. Moreover, (iii) I could not find a better
example than the last part of your comment to illustrate why ``decoherence"
and ``consistency" should be allowed to keep their original meanings, so that
we can avoid further linguistic misunderstandings and focus on the physics
instead.

\end